# Feature-interactive Siamese graph encoder-based image analysis to predict STAS from histopathology images in lung cancer

Liangrui Pan[1+], Qingchun Liang[2,3+], Wenwu Zeng[1], Yijun Peng[1], Zhenyu Zhao[4], Yiyi Liang[5], Jiadi Luo[2,3*], Xiang Wang[4*] Shaoliang Peng[1*]

[1] College of Computer Science and Electronic Engineering, Hunan University, Changsha 410082, China

[2] Department of Pathology, The Second Xiangya Hospital, Central South University, Changsha, 410011, Hunan, China

[3] Hunan Clinical Medical Research Center for Cancer Pathogenic Genes Testing and Diagnosis, Changsha, Hunan, 410011, China

[4] Department of Thoracic Surgery, The Second Xiangya Hospital, Central South University, Changsha, 410011, Hunan, China

[5] Oncology Department and State Key Laboratory of Systems Medicine for Cancer of Shanghai Cancer Institute, Renji Hospital, School of Medicine, Shanghai Jiaotong University, Shanghai, 200127, China

*To whom correspondence should be addressed. jiadiluo@csu.edu.cn, wangxiang@csu.edu.cn or slpeng@hnu.edu.cn

## ABSTRACT

Spread through air spaces (STAS) is a distinct invasion pattern in lung cancer, crucial for prognosis assessment and guiding surgical decisions. Histopathology is the gold standard for STAS detection, yet traditional methods are subjective, time-consuming, and prone to misdiagnosis, limiting large-scale applications. We present VERN, an image analysis model utilizing a feature-interactive Siamese graph encoder to predict STAS from lung cancer histopathological images. VERN captures spatial topological features with feature sharing and skip connections to enhance model training. Using 1,546 histopathology slides, we built a large single-cohort STAS lung cancer dataset. VERN achieved an AUC of 0.9215 in internal validation and AUCs of 0.8275 and 0.8829 in frozen and paraffin-embedded test sections, respectively, demonstrating clinical-grade performance. Validated on a single-cohort and three external datasets, VERN showed robust predictive performance and generalizability, providing an open platform (http://plr.20210706.xyz:5000/) to enhance STAS diagnosis efficiency and accuracy.

## INTRODUCTION

Spread through air spaces (STAS) in lung cancer was first discovered and described in 2013. Onozato and colleagues, using three-dimensional reconstruction techniques, identified islands of tumor cells outside the main tumor body's borders in the lung parenchyma, a phenomenon closely associated with tumor recurrence in patients [1]. In lung adenocarcinoma, Kadota and colleagues reported a pathological phenomenon of tumor STAS defined as the spread of lung tumor cells through the alveolar spaces to the adjacent lung parenchyma [2]. Subsequently, the World Health Organization (WHO) categorized the STAS as a pattern of tumor invasiveness in its classification of lung adenocarcinoma. [3]. STAS consists of micropapillary clusters, solid nests, or single cancer cells that invade the surrounding lung parenchyma's air spaces [4]. Research groups worldwide have published data on over 3,500 patients, finding that 15% to 69% of lung adenocarcinoma patients exhibit STAS, strongly correlating with lower survival rates and higher recurrence [5], [6], [7], [8], [9], [10], [11]. Specifically, Hassan A Khalil's team analyzed the pathology and clinical characteristics of 787 lung cancer surgical specimens and found that overall survival and recurrence-free survival were significantly lower in the STAS group compared to the non-STAS group, while the incidences of locoregional and distant recurrence nearly doubled. However, quantifiable measures of STAS did not appear to correlate with recurrence or survival metrics [10]. The International Association for the Study of Lung Cancer (IASLC) Staging Project for Lung Cancer recommended introducing STAS as a histologic descriptor in the Ninth Edition of the TNM Classification of Lung Cancer, after analyzing 4061 Pathologic Stage I NSCLC cases [12]. Moreover, clinical studies demonstrated that limited resection was associated with worse survival than lobectomy in stage I STAS patients [13]. Thus, accurately identifying STAS is of great significance for assessing prognosis and guiding surgical extent in clinical application.

Histopathology is currently the gold standard for diagnosing STAS in lung cancer [14], [15]. However, pathologists' assessment of STAS is subjective, prone to missed or misdiagnoses, time-consuming, and labor-intensive, making it unsuitable for large-scale histopathological diagnostics [16]. Intraoperative frozen section (FS) diagnosis assists clinicians in making decisions during surgery, including adjusting the scope and methods of surgery and promptly assessing surgical margins. Clinical studies have shown that lung cancer patients with STAS in stage T1 might have better survival outcomes with lobectomy compared to sublobar resection [5]. Paraffin-embedded sections (PSs) are also a fundamental method in pathological examinations, used to study the pathogenesis, pathophysiology, and molecular biology characteristics of diseases, providing a scientific basis for disease prevention, treatment, and control [17][18]. Yun et al. reported that the accuracy of FS diagnosis for STAS was 74.14%, with a sensitivity of 55.14%, a specificity of 85.02%, and moderate agreement [19]. Similarly, a large multi-center prospective observational study evaluating STAS in FS of patients with cT1N0M0 invasive lung adenocarcinoma found that FS was highly specific but lacked sensitivity for STAS detection, with moderate agreement [20]. However, the current accuracy of diagnosing STAS using FSs and PSs remains low, with high rates of false positives and false negatives, and the workload is heavy, making the use of AI to predict STAS on FSs and PSs an urgent need [19], [21], [22].

Benefiting from the rapid development of artificial intelligence, deep learning has been widely applied to tasks in image analysis, providing state-of-the-art (SOTA) performance for various image classification tasks [23], [24]. In computational pathology, deep learning-based image analysis has demonstrated performance comparable to that of pathologists across various tasks. For instance, deep learning enabled semi-supervised training to classify prostate histopathology images and generalize to entirely different datasets [25]. Deep learning-based classification of liver cancer histopathology images achieved an accuracy of 96%, nearing the capability of pathologists with five years of experience [26]. Recent studies have shown that attention-based Transformer can perform global feature extraction and allow interactions among features at different positions, enhancing the model's ability to capture relationships across various image regions, thus improving classification performance. For example, a multi-instance learning (MIL) network based on the Swin Transformer using whole slide image (WSI) labels accurately classified colorectal cancer WSIs [27]. A breast cancer histopathology image classification model using color deconvolution and the Transformer architecture reached an average precision of 93.02% [28].

Considering the proximity of most STAS to the main tumor, constructing a spatial topological map of WSI is essential for STAS diagnosis. The spatial topological map represents each independent patch as a point, using edges between points to define relationships among patches. It integrates global structural information into the graph, enabling the model to consider WSI characteristics comprehensively, rather than focusing solely on localized areas [29]. Additionally, the spatial topological map visualizes spatial relationships and structural characteristics across different WSI regions, enhancing the interpretability and understandability of model classification results [30]. The WSI spatial topological map also reflects the tumor's microenvironment, crucial for cancer research, diagnosis, and treatment [31]. For WSIs with ultra-large pixels, constructing their spatial topological map enhances data processing and boosts the reasoning efficiency of neural networks.

Here, we propose for the first time that the diagnosis of STAS can be predicted through image analysis based on a feature-interactive Siamese graph encoder (VERN). To our knowledge, this is the inaugural study to forecast STAS using WSI-level labels. Initially, we constructed a large, single-cohort histopathology image dataset of STAS lung cancer, consisting of 1546 FSs and PSs, for internal validation and testing via VERN. In the internal test set, VERN attained an area under the curve (AUC) of 0.9215. Crucially, the VERN accurately predicts STAS in both intraoperative FSs and PSs, matching clinical diagnostic outcomes. VERN calculates attention contribution values for each patch and visualizes WSIs, aiding pathologists in focusing on high-risk and peripheral tumor patches, thus enhancing diagnostic accuracy and efficiency. VERN was extensively tested on STAS validation sets at three additional centers, confirming the model's effectiveness and generalizability. Additionally, we investigated survival and recurrence in non-STAS and STAS patient cohorts based on TNM staging and analyzed lung cancer-related protein expressions (PD-L1, P53, Ki67, ALK). Statistical analyses reveal a strong correlation between STAS, cancer staging, and protein expression levels. Last but not least, we first established an open STAS prediction website(http://plr.20210706.xyz:5000/), enabling pathologists to obtain predictions results by uploading WSIs.

## Results

### Feature-interactive Siamese graph encoder-based STAS prediction outperforms the SOTA

We utilized the VERN to predict STAS in a dataset of lung cancer patient histopathology (Figure 1). Initially, we applied a five-fold cross-validation method to divide the internal validation set, which includes both FSs and PSs, into five subsets. We used four subsets for training during each phase, with the remaining subset reserved for testing. This approach yielded the five best STAS prediction models. We observed that the model from the third fold of cross-validation achieved a total area under the receiver operating characteristic (AUCROC) curve of 0.9215 in domain-specific tests (Figure 2a), while the fourth fold model achieved a precision-recall curve (PRC) of 0.9491 (Figure 2b). The average in-domain test AUROC for the five-fold cross-validation model was 0.8683 (see Figure 2a), and the average in-domain test PRC was 0.9122 (Figure 2b). Among them, the predictive results of VERN after five-fold cross-validation training were expressed using mean and standard deviation, and the accuracy, precision, recall, F1 score, specificity, and AUC values were approximately 0.7926, 0.7828, 0.7899, 0.7845, 0.7795, and 0.8683, respectively (Figure 2c). The confusion matrix (Figure 2d) indicated that VERN has achieved high performance in predicting STAS using five-fold cross-validation. Lastly, we have trained ABMIL, DSMIL, TransMIL, DTFD-MIL, and interventional bag multiple instance learning (IBMIL) models using the five-fold cross-validation method based on the Second Xiangya Hospital dataset [32], [33], [34], [35], [36]. We evaluated model performance using metrics such as accuracy, precision, recall, F1-score, specificity, and AUC, as shown in Table 1. Analysis of Table 1 data shows that VERN accurately predicted the presence or absence of STAS in pathological images, outperforming current SOTA methods. This outcome may be attributed to the construction of the spatial topology of histopathological images based on features extracted by KimiaNet and CTransPath, and improved prediction of STAS's potential labels through the inference of the VERN. It is noteworthy that the VERN trained by the fifth fold of cross-validation exhibited a prediction bias, possibly due to category imbalance in the training and testing data, which made it challenging for the VERN to learn effective feature representations during training and validation. Additionally, there is some degree of randomness in the division of data and the initialization of the model during the cross-validation process.

### VERN predictive performance exceeds clinical grade

To validate the effectiveness of the VERN, we conducted tests using an internal testing dataset that included 40 FSs and 100 PSs. Extensive literature indicates that, compared to PSs, FSs have a lower sensitivity for detecting STAS, with clinical diagnostic accuracy at 74% [19], [22] and an AUCROC of 0.67 [21]. However, in our experiments, the VERN achieved an AUROC of 0.8275 on FSs and 0.8454 for PRC in domain-specific tests. For PSs, the VERN's in-domain AUROC was 0.8829, and the PRC was 0.9165. We observed that the VERN performed less effectively in predicting STAS in FSs, potentially due to their limited involvement in model training. Additionally, the quality of FSs may be inferior, with damaged cellular structures and fewer samples, which could lead to significant errors in the model's judgments. However, by examining the confusion matrix of the VERN for FSs, we found an overall accuracy of 80%. In contrast, in PSs, the VERN's overall prediction accuracy reached 85%, with higher precision in predicting STAS on WSI. This improvement is likely due to the consistent quality of PSs, intact cellular structures, and extensive representation in the internal dataset, which aided in constructing spatial topological maps, enabling effective STAS inference by the VERN.

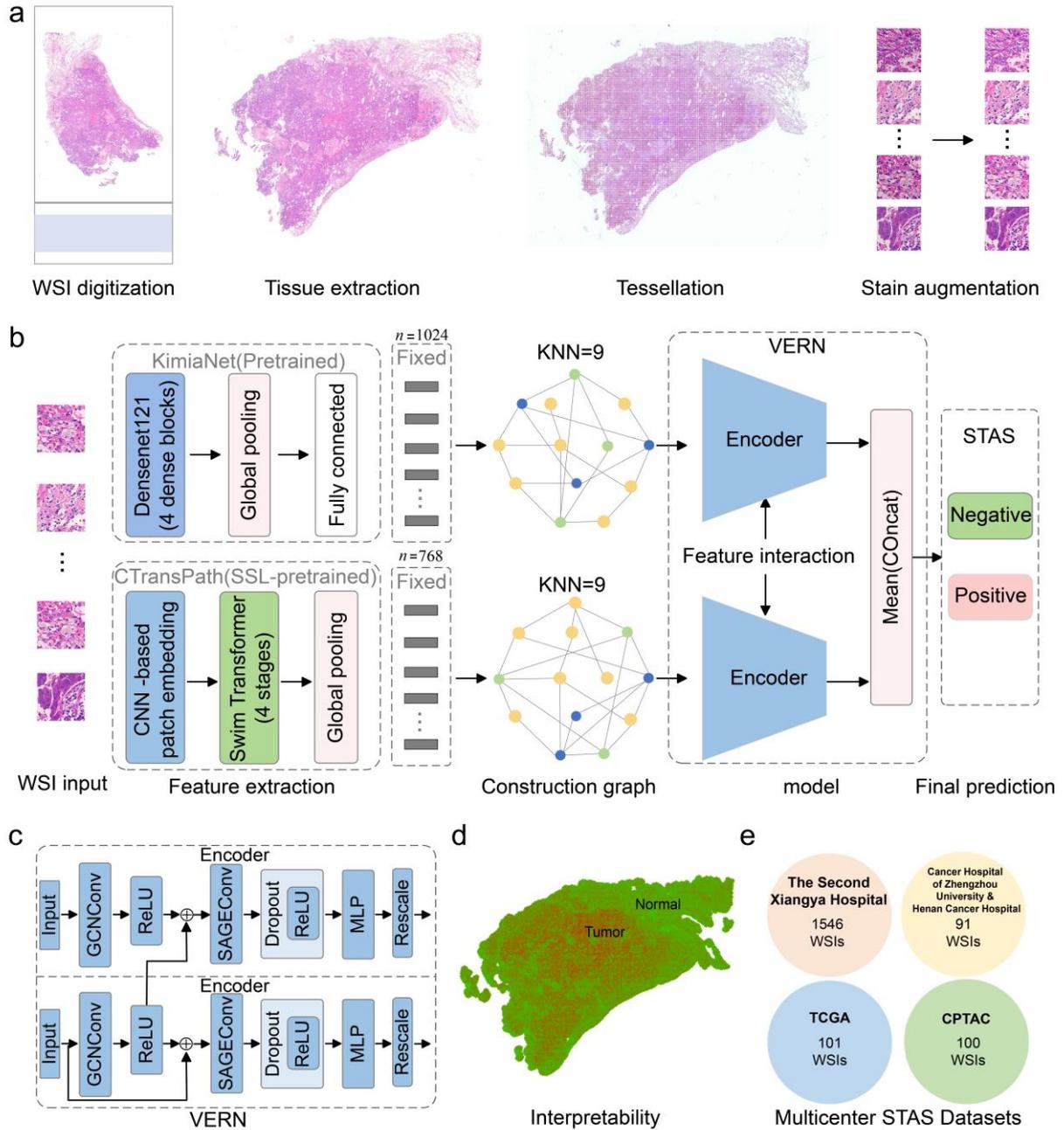

Figure 1. Workflow including data preprocessing, model training and inference, interpretability analysis, and multicenter validation. **a** Data preprocessing involves digitization of WSI, extraction of regions of interest, segmentation of tissue into patches, and patch data augmentation. **b** The model architecture includes feature extraction from pretrained models (KimiaNet, CTransPath), construction of WSI spatial topological maps, feature-interactive Siamese graph encoder (VERN) module, and diagnostic results for STAS. **c** Detailed architecture of the VERN. **d** Interpretability analysis of WSI. **e** The STAS dataset includes internal training and validation sets, a test set, and an external validation set.

Table 1. The multiple instance learning methods predict the SOTA result of STAS.

|  |  | Accuracy | Precision | Recall | F1-score | Specificity | AUC |
|---|---|---|---|---|---|---|---|
|  | ABMIL | 0.6905 | 0.7582 | 0.6711 | 0.7048 | 0.5404 | 0.7691 |
|  | DSMIL | 0.5529 | 0.6189 | 0.6691 | 0.6359 | 0.7391 | 0.6489 |
|  | TransMIL | 0.6878 | 0.6981 | 0.7373 | 0.7172 | 0.6957 | 0.7619 |
|  | DTFD-MIL | 0.6825 | 0.6865 | 0.7041 | 0.6951 | 0.6708 | 0.769 |
| IBMIL | ABMIL | 0.7169 | 0.7177 | 0.7712 | 0.7186 | 0.7143 | 0.7768 |
|  | DSMIL | 0.5952 | 0.6744 | 0.6347 | 0.6526 | 0.6335 | 0.707 |
|  | TransMIL | 0.7011 | 0.7026 | 0.7156 | 0.709 | 0.6708 | 0.7394 |
|  | DTFD-MIL | 0.7222 | 0.7197 | 0.7084 | 0.7119 | 0.6149 | 0.7718 |
|  | VERN | 0.7926 | 0.7828 | 0.7899 | 0.7845 | 0.7795 | 0.8683 |

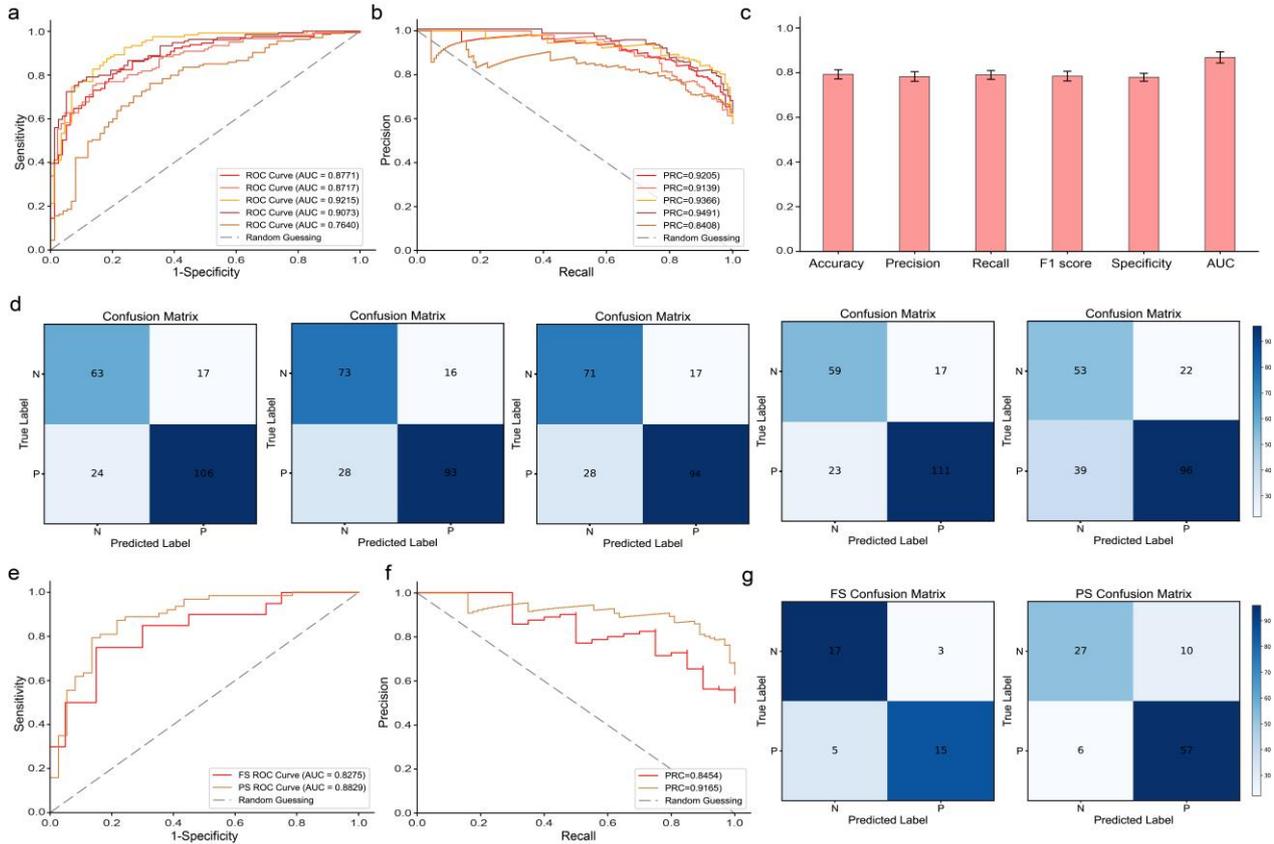

Figure 2. Experimental results of STAS prediction using the VERN trained with five-fold cross-validation on internal and external datasets. **a** ROC curves of the VERN from five-fold cross-validation. The five curves represent the testing performance of the VERN after each fold of training, with the diagonal line showing the in-domain test results. **b** PRC of the VERN from five-fold cross-validation. **c** Accuracy, precision, recall, F1 score, specificity, and AUC values of the VERN from five-fold cross-validation testing. **d** Confusion matrix of the VERN from five-fold cross-validation testing. **e** ROC curves of the VERN tested on digital FSs and PSs. **f** PRC curves of the VERN tested on digital FSs and PSs. **g** Confusion matrix results of the VERN tested on digital FSs and PSs.

**Feature-interactive Siamese graph encoder-based workflows are explainable**

In this manuscript, we aim to build an interpretable VERN, helping researchers and clinicians better understand the model's decision-making. Heatmaps visually display regions predicted by VERN in WSI images, based on each patch's contribution (probability of tumor vs. non-tumor), enabling intuitive understanding of which features impact the model's predictions. This helps clinicians quickly identify areas of interest, reducing the need for manual inspection and enhancing diagnostic efficiency. WSI heatmaps also highlight potential abnormal areas, helping clinicians detect overlooked lesions. We first require pathologists to annotate the main tumor areas and STAS zones in the WSI (Figure 3a). In non-STAS images, only the main tumor body is delineated. In STAS-containing images, micropapillary clusters, solid nests, and single-cell type STAS are prominently distributed around the main tumor area, as shown in magnified regions. We then visualize VERN's final contributions for each patch and map them back to the WSI based on location to assess their utility in recognizing tumor areas (Figure 3b). Some patches may be misclassified as tumor or non-tumor due to tissue heterogeneity or segmentation strategies, leading to anomalous contribution levels.

In non-STAS WSIs, patches with higher contributions are mainly concentrated in the main tumor areas. In STAS-containing WSIs, highly contributive patches are distributed in the main tumor areas and extend around the tumor body. These color-coded markings help pathologists identify areas of interest through overall observation. Patches with lower contributions are located in the stromal and normal tissue areas. As STAS often appears at the edges of the tumor, visualized WSIs guide pathologists to focus on these areas, saving time and improving efficiency. Heatmaps cannot directly locate STAS because patches containing STAS also include other cells, and their contributions are lower than those containing mainly tumor cells. The nine highest-contributing patches identified by VERN (Figure 3b) are mainly located in the main tumor body, aiding differentiation between pathological and normal tissues and improving VERN's interpretability. Pathological features in high-interest areas correlate with STAS types; micropapillary STAS is histologically similar to the main tumor, containing many micropapillary structures. The other two STAS types do not show clear consistency with the main tumor's histology. Furthermore, they provide cellular-level features, including nuclear morphology and mitotic activity, offering significant data support for studying tumor growth patterns and metastatic characteristics [37].

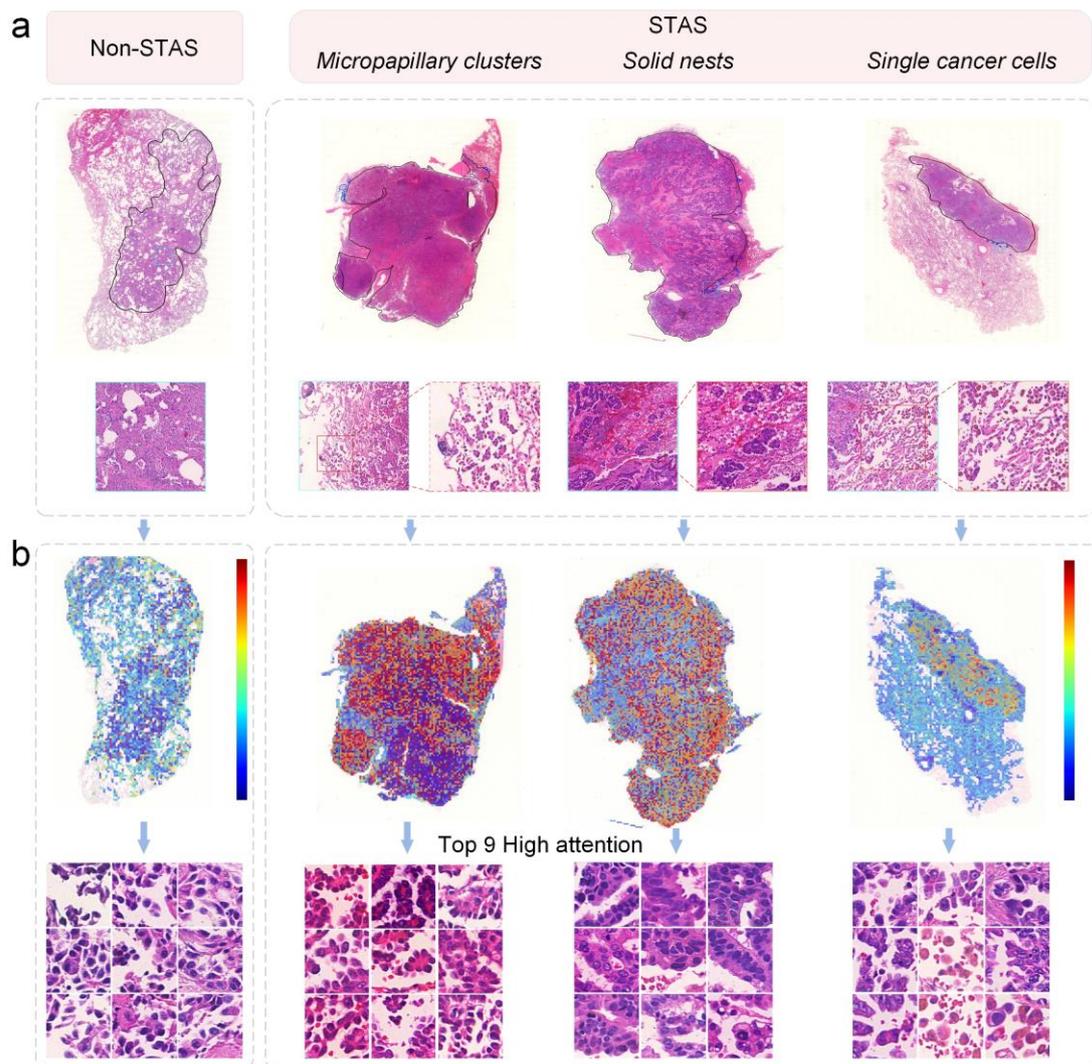

Figure 3. Interpretability analysis of the VERN. **a** WSIs with STAS-negative and STAS-positive annotations, where pathologists outlined the main tumor body and STAS subtypes, including micropapillary, solid nests, and single cells. **b** Attention scores for each patch based on the VERN. High values (red) indicate higher model-predicted contributions, while low values (purple) indicate lower contributions. Additionally, the nine patches with the highest contributions were identified.

**Single-cohort and multicenter STAS validation sets validated the generalizability of the VERN**

We selected 356 histopathological images from patients who underwent lung nodule resection and were diagnosed with lung cancer at the Second Xiangya Hospital of Central South University. Among them, each patient took a histopathological image that had not been internally trained, verified, and tested. We used the VERN to predict STAS in these 356 histopathological images. The VERN achieved an AUROC of 0.9181 and a PRC of 0.941 in in-domain tests (Figure 4a). Other metrics, including accuracy, precision, recall, F1 score, specificity, and AUC (Figure 4e), all exceeded 0.8, demonstrating the model's accuracy and practicality in predicting STAS. Additionally, we validated the VERN using 91 pathological sections from the Cancer Hospital of Zhengzhou University and Henan Cancer Hospital. The VERN achieved an AUROC of 0.8699 and a PRC of 0.7545 in intra-domain tests (Figure 4b). Except for specificity, all other evaluation metrics (Figure 4e) exceeded 0.76, confirming the model's good generalization ability, comparable to experienced pathologists.

Finally, we utilized lung cancer histopathological images from TCGA and CPTAC to construct two external validation sets to assess the model's validity and generalization performance. In the TCGA test set, the VERN achieved an in-domain test AUROC of 0.7029 and a PRC of 0.6225 (Figure 4c). Other evaluation metrics, including accuracy, precision, recall, F1 score, and AUC (Figure 4e), all exceeded 0.75. In the CPTAC test set, the VERN achieved an in-domain test AUROC of 0.7555 and a PRC of 0.8025 (Figure 4d). Other evaluation metrics (Figure 4e) all exceeded 0.76. The performance of the VERN is comparable to that of intermediate to senior pathologists.

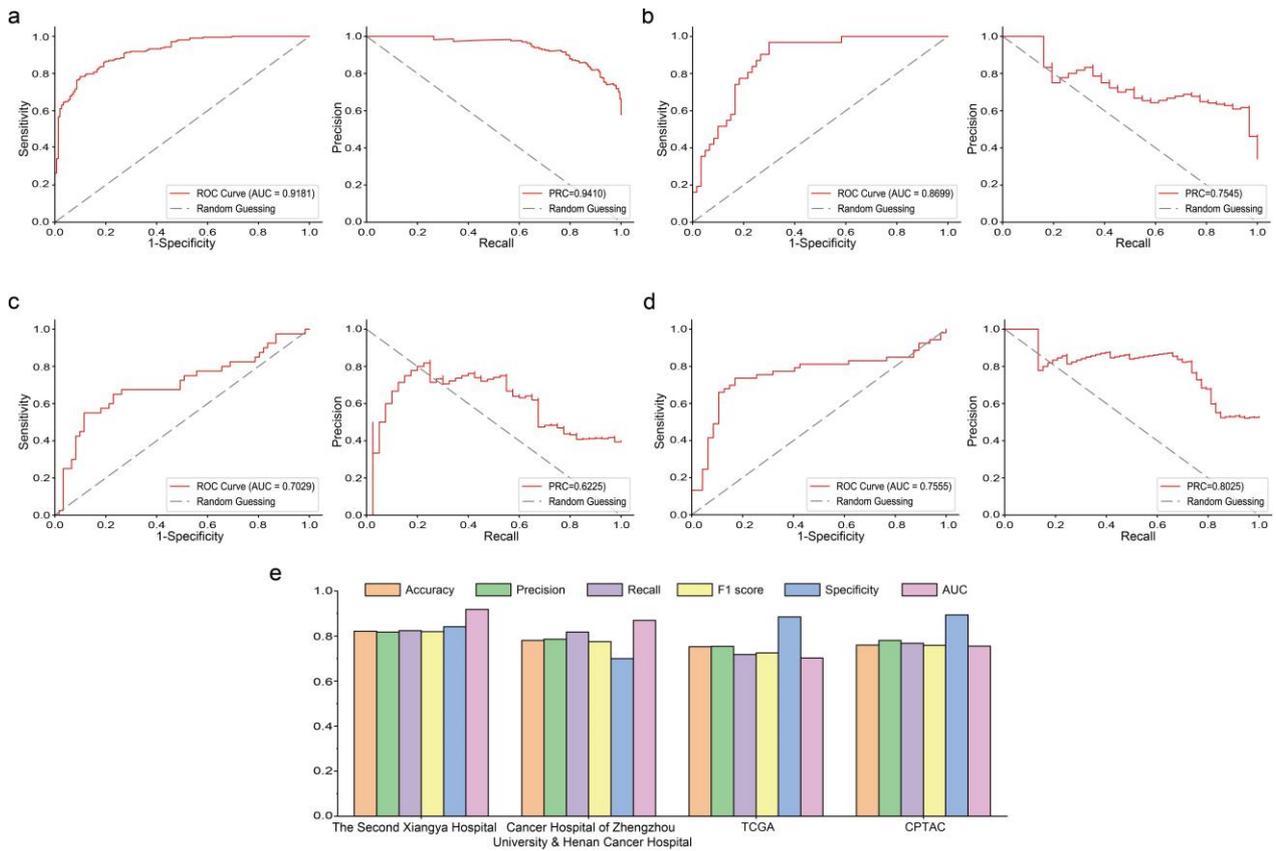

Figure 4. Evaluation of VERN performance in predicting STAS in single-cohort and multicenter experiments. **a** ROC and PRC curves validating the VERN's effectiveness based on 356 histopathological images from the Second Xiangya Hospital of Central South University. **b** ROC and PRC curves validating the VERN's effectiveness based on 91 histopathological images from the Cancer Hospital of Zhengzhou University and Henan Cancer Hospital. **c** ROC and PRC curves validating the VERN's effectiveness based on 101 histopathological images from TCGA. **d** ROC and PRC curves validating the VERN's effectiveness based on 100 histopathological images from CPTAC. **e** Accuracy, precision, recall, F1 score, specificity, and AUC values of the VERN predicting the presence or absence of STAS across the four datasets.

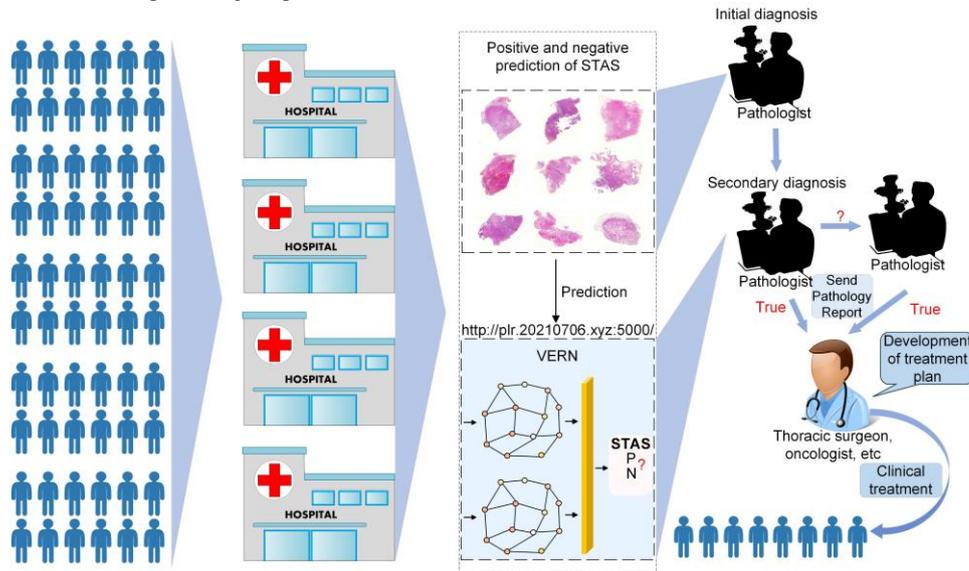

Figure 5. Workflow diagram of the application of the VERN in clinical practice. This system can be used as a supplement to the interpretation of STAS by pathologists or pathology departments in hospitals in areas with insufficient medical resources. The process involves patients undergoing surgery in the hospital, pathology sections being prepared and scanned by the pathology department, and testing through the STAS testing website. The three-level diagnosis and treatment system is combined with the test results of the VERN. The pathologist provides a detailed pathological diagnosis report, and the thoracic surgeon determines the surgical method suitable for the patient (such as lobectomy, sublobar resection, etc.), and the oncologist develops a follow-up treatment plan to provide patients with personalized and precise treatment. We welcome you to visit our test platform http://plr.20210706.xyz:5000/, or download the VERN from https://github.com/pengsl-lab/STAS/tree/main to use.

**Feature-interactive Siamese graph encoder-based workflow enables clinical application**

The status of STAS in lung cancer patients needs to be clarified through histopathology. Early studies have shown that the surgical approach and survival outcomes of STAS patients are closely related [4]. Therefore, in the clinical setting, determining the presence of STAS via intraoperative FSs is crucial for choosing the appropriate surgical method. Our proposed VERN achieved high accuracy in detecting STAS in both FSs and PSs, surpassing other models reported in the literature [19], [21], [22]. In practical clinical settings, pathologists face low accuracy and poor repeatability in interpreting STAS from lung cancer FSs, while PSs interpretation, though more accurate, requires more time and effort from pathologists, and the heavy clinical workload negatively impacts the precise interpretation of lung cancer STAS. Our VERN can effectively assist in predicting STAS, saving manpower and reducing the workload for pathologists, allowing patients to receive precise surgical interventions tailored to their conditions. Particularly in regions with limited medical resources, where the level of medical care is constrained, the VERN can help improve the quality of diagnostic pathology and surgical standards. Additionally, AI-based WSI analysis can also help pathologists improve their diagnostic skills for STAS, alleviating the challenges brought by the uneven distribution of medical resources. Therefore, employing the VERN to guide STAS prediction is urgent.

We proposed a specific workflow suitable for clinical practice (Figure 5). Initially, patients with lung nodules undergo biochemical and imaging examinations within the hospital to confirm surgical indications. Patients with indications then have their surgery timing and planned approach confirmed. During the surgery, the surgeon removes the lung nodule and surrounding lung tissue, which are immediately sent to pathology for intraoperative sampling and the preparation of FSs, followed by scanning to obtain high-resolution histopathological images for diagnosis by pathologists. Using the VERN to assist in predicting STAS in WSIs can reduce the time pathologists spend reviewing slides, thereby reducing the intraoperative waiting time for patients, benefiting them. Clinical doctors further guide the extent of surgical resection and lymph node dissection based on the intraoperative FS results provided by pathologists in conjunction with the VERN.

## Discussion

Deep learning and attention mechanisms are extensively used for diagnosing tissue WSI. However, for accurate STAS diagnosis, relying solely on WSI features is insufficient; spatial graphs are also necessary to capture the locational relationships within cancer tissues. Most deep learning approaches like CNNs, LSTMs, GANs, and VAEs primarily capture local positional information from each patch [38]. While attention mechanisms provide some level of global positional encoding, they typically focus on adjacent or within-a-certain-distance features. In contrast, spatial topological maps cover a broader spatial range, enabling models to capture relationships from more distant locations. Therefore, representing WSI features with graphs is essential for STAS prediction tasks.

In this study, we introduced the VERN for predicting STAS in FSs and PSs of lung cancer. This model comprises feature-interactive Siamese graph encoder for extracting spatial topological features from WSI and a feature aggregation module that updates node features and aggregates edge characteristics to ascertain the positional relationships between STAS and the primary tumor mass. The VERN demonstrated an AUCROC of 0.8275, a PRC of 0.8454, and 80% accuracy on FSs. On PSs, it achieved an AUCROC of 0.8829, a PRC of 0.9165, and 85% accuracy. These outcomes may be attributed to the model's robustness, enhanced by training on both FSs and PSs. Additionally, VERN outperformed existing methods in feature extraction and classification, suggesting its potential as an accurate tool for predicting STAS, providing valuable insights for pathologists and clinicians.

In the internal validation set at the Second Xiangya Hospital of Central South University, VERN's accuracy in detecting STAS on FS was lower than on PS, likely due to the inherent characteristics of FS. Literature indicates that multiple senior pathologists assessed STAS in FS of lung cancer specimens, using PS as the gold standard. Detection of STAS on FS was suboptimal, with accuracy between 74.14%-85.1%, sensitivity of 55.14%-56.4%, and specificity of 85.02%-95.4% [19], [20]. Sensitivity in STAS interpretation on FS is low, with an accuracy of approximately 80% under pathologist review. VERN, as an AI-based model, derives its prior knowledge from rigorously interpreted pathology slides (both FS and PS) by two senior pathologists, followed by parameter tuning. Therefore, VERN's lower detection accuracy of STAS on FS compared to PS is consistent with previous findings. The lower detection rate of STAS on FS by VERN may also stem from several factors: FS sampling is limited to a single tissue section, reducing representativeness, while PS generally includes the entire tumor, providing a more comprehensive STAS assessment. Additionally, ice crystals often form during FS preparation, impacting data quality and interpretability, complicating FS interpretation. Poor intraoperative FS quality is often due to technical defects (e.g., thick slices, deformed cells, ice crystals, cracks, and light HE staining) and sampling issues (e.g., insufficient tissue or improper selection). Consequently, data imbalance and quality issues contribute to FS's lower diagnostic accuracy compared to PS. From the VERN model's perspective, the pathology dataset at the Second Xiangya Hospital of Central South University includes both FS and PS samples, and VERN was trained using five-fold cross-validation. With fewer FS samples compared to PS, VERN acquired more prior knowledge from PS, leading to lower diagnostic performance on FS during testing. Interestingly, a re-evaluation of FSs indicated that 61.3% of STAS cases were initially misdiagnosed as non-STAS [39]. Pathologists should communicate the low sensitivity of FSs for STAS to surgeons and reconsider the protocols for preparing these sections to ensure they adequately include neighboring lung parenchyma and tumors. Additionally, knife-spread artifacts, occurring during the slicing and handling of lung tissues, may lead to artificial displacement of tumor cells and surrounding parenchyma, potentially transferring loose tissue fragments into airways, bronchioles, or alveoli, and confounding them with STAS [40]. Thus, Standardizing FS sampling, optimizing preparation processes, and reducing artifacts can improve VERN's detection accuracy for STAS in clinical practice.

The VERN faces challenges in interpretability and increased complexity in classification tasks. Graph structures often have multiple layers with numerous nodes and edges, leading to increased model complexity as the network deepens, complicating the understanding of each node and edge's role [41]. Significant errors are prevalent when projecting node feature scores back onto histopathological images. Images with high or low feature scores show distinct characteristics. Compared to non-STAS histopathological images, STAS typically appears near the primary tumor with three main unevenly distributed pathological features. Pathologists can use heat maps to focus on areas with higher or lower feature scores to predict STAS. AI-based feature capture also aids physicians in identifying highly expressed tumor areas, potentially linked to the tumor's etiology.

In multicenter validation tests, the VERN achieved a prediction accuracy of 82.17% in the internal validation (the Second Xiangya Hospital of Central South University) and 78.02% in the validation (Cancer Hospital of Zhengzhou University and Henan Cancer Hospital). The relatively high accuracy was primarily due to the superior quality of pathological slides and high-resolution images from advanced scanners. In Cancer Hospital of Zhengzhou University and Henan Cancer Hospital dataset, VERN identified 18 false positives and two false negatives, with false positives often due to misinterpretations of lymphoid aggregates or detached bronchial epithelial cells around the primary tumor. Notably, the model generalizes well to high-quality external test data. However, the VERN model has not yet acquired any prior knowledge from external multi-center data, leading to a decrease in diagnostic accuracy. In TCGA, VERN predicted STAS with 75.25% accuracy, including 7 false positives and 18 false negatives, likely due to poor image quality and misidentification of deeply colored tissue cells, lymphocytes, or stroma as cancerous. The proportion of micropapillary disseminated foci was small, and VERN failed to learn from a large number of WSI containing micropapillary clusters features, leading to unrecognized micropapillary STAS during testing. In CPTAC, VERN achieved a 76% accuracy rate, with 19 false negatives and five false positives, mainly due to misjudgments of lymphoid aggregates or carbon dust around the primary tumor and non-recognition of micropapillary disseminated foci, along with some poor-quality slides.

STAS detection serves a direct clinical function. It is well known that clinical staging is a key factor in determining patient prognosis (Supplementary Figure 2). Through our cohort at the Second Xiangya Hospital of Central South University, we observed STAS in very early stages of lung cancer (Supplementary Figure 1a), underscoring its aggressive nature, consistent with our findings that STAS patients are more prone to pleural, vascular, and neural involvement (Supplementary Figure 3h). The formation of high aggressiveness may relate to two factors: the lung's unique physiological structure, which is an organ composed of the bronchial tree and numerous alveoli that continually facilitate gas exchange through respiratory movement. When a tumor occurs, the main body of the tumor also moves passively with breathing, making it easier for cancer cells to detach and spread distantly [42]. On the other hand, it is related to the type of lung cancer itself. Lung adenocarcinoma has become the most prevalent type of lung cancer (Supplementary Figure 2g), which includes mucinous, solid, papillary, acinar, lepidic, and micropapillary adenocarcinoma [43]. Typically, a lung adenocarcinoma patient has multiple tissue components, with mucinous components likely to move into the airways with mucus flow, and micropapillary components prone to detachment, thus enhancing their invasiveness. Our data faithfully reflect that patients in the STAS group, with higher clinical stages, are significantly more than those in the non-STAS group. Patients in the STAS group are more likely to experience lymph node metastasis, distant metastasis, and are more prone to recurrence and have a worse prognosis (Supplementary Figure 1), consistent with literature on STAS research [3], [44], [45], [46], [47].

PD-L1, P53, Ki67, and ALK are key proteins related to the prognosis and treatment of lung cancer. Studies show a high concordance between P53 protein expression and TP53 gene mutations, indicating that P53 protein levels can accurately predict TP53 mutations [48], [49]. Similarly, strong expression of ALK protein precisely indicates ALK gene rearrangement [50]. Our clinicopathological analysis found that STAS patients have significantly higher levels of PD-L1, P53, Ki67, and ALK compared to the non-STAS group (Supplementary Figure. 3), aligning with most literature [51], [52], [53], [54]. However, Arne Warth's team in a retrospective study of lung adenocarcinoma noted no difference in the Ki67 proliferation index between STAS and non-STAS groups [3], which may relate to their method of assessing Ki67. They compared median values of Ki67 proliferation indices between groups, whereas our evaluation uses categorical expressions of Ki67, reducing systematic errors for more objective conclusions [55]. Our findings suggest that mutations in the tumor suppressor gene P53, higher Ki67 indices, and ALK gene rearrangements predispose to STAS, affirming its high invasiveness at the protein and gene levels. Fortunately, the increased expression of PD-L1 and the higher incidence of ALK rearrangements in STAS offer more opportunities for immunotherapy and targeted treatments for these patients.

An interesting finding from our study is that STAS was not only observed in lung adenocarcinoma and squamous cell carcinoma but also in small cell lung cancer and large cell neuroendocrine carcinoma (Supplementary Table 2), and even in metastatic lung cancers, indicating that STAS is not exclusive to primary lung tumors. Our single-center lung cancer cohort showed the highest frequency of STAS in predominantly acinar adenocarcinoma, followed by solid and then papillary subtypes. These data could further elucidate the causes and mechanisms of STAS occurrence.

STAS exhibits high invasiveness in lung cancer and is rarely considered by clinicians during wedge resection surgeries. Since the WHO's formal recognition of STAS in 2015, our study is the first to construct a STAS dataset based on 1546 histopathological images. Importantly, we are also the pioneers in building spatial topological maps and predicting STAS using the VERN, which demonstrated AUROC scores of 0.8275 for FSs and 0.8829 for PSs. The effectiveness and generalizability of the VERN were confirmed by multicenter STAS validation sets. Although the diagnosis of STAS is made by VERN predicting each patient's WSI, if one of the WSIs is diagnosed with STAS, the patient can preliminarily be considered as having STAS. Given its smaller parameter count, graph neural networks can be effectively deployed in end-service centers. We offer an open website for STAS prediction and plan continuous updates to provide the best prediction models, enabling researchers and clinicians to automatically predict STAS in pathological sections. This tool assists physicians in devising optimal, personalized treatment plans based on patients' STAS prediction and cancer tissue protein expression profiles.

## Methods

### Clinical single cohort and multicenter data collection

From April 2020 to December 2023, we randomly selected 356 patients who underwent lung nodule resection surgery and were diagnosed with lung cancer at the Second Xiangya Hospital to form a retrospective lung cancer study cohort. Each patient's tumor specimen was prepared into FSs and multiple PSs, along with several immunohistochemical slides for treatment and prognostic indicators. Comprehensive clinicopathological data such as age, tumor size, lymph node metastasis, distant metastasis, clinical staging, recurrence, and survival were collected. Two pathologists reviewed all patients' pathology data to confirm the presence of STAS, the specific pathology types of disseminated foci, the specific pathology types of lung cancer, and the expression of target proteins (PD-L1, P53, Ki-67, ALK). Of these, 150 patients were non-STAS, and 206 had STAS. A total of 1546 histopathological images were collected and scanned, of which 1190 (213 FSs, 977 PSs) were divided into groups for internal validation (173 FSs, 877 PSs) and testing (40 FSs, 100 PSs) of model performance. Another 356 PSs from the single-cohort not included in internal training and testing were used for multicenter validation of model performance.

Based on our research objectives, only WSIs of patients meeting the following criteria were included in the study: 1) confirmed diagnosis of LUAD, 2) corresponding routine pathological slides, including the main tumor body and adjacent non-tumor tissue, 3) detailed TNM staging, 4) high-quality slides, such as those without bending, wrinkling, blurring, or color changes, 5) excluding slides that contain the main tumor body but lack adjacent non-tumor tissue. The external validation set included pathological images from three centers. Two experienced pathologists, using microscopes, labeled each WSI with STAS following a double-blind principle and cross-validation to ensure accuracy and reduce subjectivity, missed, or over-diagnosis. The Cancer Hospital of Zhengzhou University and Henan Cancer Hospital provided 91 digital pathology slides (60 STAS and 31 non-STAS) that were used to create an STAS validation set. Importantly, we also collected 101 digital pathology slides from TCGA (40 STAS and 61 non-STAS) and 100 from CPTAC (53 STAS and 47 non-STAS) to build STAS validation sets for testing the model's generalization performance. Supplementary Figures 4a-b show the detailed inclusion and exclusion processes of the TCGA and CPTAC datasets.

**Methods description**

We propose a STAS prediction workflow primarily consisting of four steps: (1) Image preprocessing (Figure 1a), (2) Feature extraction based on pretrained models (KimiaNet, CTransPath), (3) Construction of spatial topological maps for the entire WSI, and (4) Feature extraction using a VERN which extracts features from the graph for final prediction (Figure 1b). Additionally, the VERN generates attention scores for patches, which are used to create visually interpretable maps and conduct downstream analysis of STAS patients.

The image preprocessing step primarily involves digitizing the WSI, using an RGB filter to segment tissue areas, and detecting background and blurry regions [56]. All WSIs are automatically processed to generate thumbnail, mask, and overview images. Subsequently, the tissue regions in the WSI at 20X magnification are segmented into 512 × 512 pixel patches, with each patch's coordinates and position recorded. To minimize the impact of dataset quality on the model's generalizability, a GAN based on pathological image features is used to enhance the features of all patches in the WSI.

To enhance the robustness and generalization of the VERN, we employ KimiaNet and CTransPath for feature extraction, yielding 1024-dimensional and 768-dimensional representations respectively for each patch, thereby enriching the diversity of feature representations within the image [57], [58]. KimiaNet, fine-tuned and trained on the DenseNet framework with a pre-trained model consisting of 7 million weights, stands apart from other convolutional neural networks (CNNs) by virtue of its ability to learn from millions of pathological patches, effectively extracting features from diverse patches [59]. The local connectivity and sliding window operation within KimiaNet's convolutional layer facilitate the capture of local features, progressively expanding the receptive field through multi-layer convolutional operations, thereby enabling hierarchical feature extraction and effective representation of input data, culminating in global pooling and fully connected output [25]. Each patch embedding is then standardized into a 1024-dimensional vector representation. Conversely, CTransPath, built upon the Swin Transformer architecture, comprises three convolutional layers and four Swin Transformer modules [60]. While the convolutional layer of CTransPath extracts local features from image data, the Swin Transformer captures global features and long-range dependencies. By amalgamating the strengths of CNNs in local feature extraction and Swin transformers in global feature capture, the model's representational capacity is enhanced. Utilizing CTransPath, we extract patch embedding features, each standardized into a 768-dimensional vector representation. Additionally, to mitigate potential model bias, we employ two pre-trained models for feature extraction in subsequent analyses.

Each patch of the WSI is represented as a feature vector. However, cells and tissue structures within the WSI exhibit relative positional relationships in a two-dimensional plane. To address this, we employ the K-nearest neighbor (KNN) algorithm (K=9) to construct a spatial topological graph ($G$) comprising nodes and edges [61], [62]. $G$ facilitates feature extraction and spatial relationship analysis within the WSI. These features may encompass cell morphology, nuclear morphology, and cell distribution. Analyzing the interrelationships among these features enhances understanding of tissue structures and pathological characteristics. In $G = \{v, e\}$, $v$ represents the node set of the graph, $e$ represents the set of edges, $A$ denotes the adjacency matrix, and $X$ signifies the node characteristic.

The VERN consists of two symmetrical encoders (Figure 1c) designed to extract features from G for binary classification. Each encoder includes a GCNConv layer, a SAGEConv layer, two ReLU layers, a Dropout(with a parameter of 0.2) layer, an Multilayer Perceptron(MLP) network and a Rescale layer in series [63], [64], [65]. Initially, the GCNConv layer initializes the convolution kernel weight matrix $W$, which multiplies the node feature $X$ and the weight matrix $W$ to obtain the linear transformation result of the node $Z = X \times W$ Z=XW. The adjacency matrix $A$ is used to propagate the characteristics of the nodes. Specifically, for any node $i$, the features of its neighboring nodes are aggregated with weights, and then added to the features of the node itself to obtain a new feature representation. The process is as follows:

$$H^{(l)} = \sigma(\hat{D}^{-\frac{1}{2}} \hat{A} \hat{D}^{-\frac{1}{2}} H^{(l-1)} W^{(l)}) \quad (1)$$

Here, $H^{(l-1)}$ represents the feature representation of the nodes at layer $l-1$, $\hat{A} = A + I$ represents the adjacency matrix $A$ augmented with the self-connection matrix, and $D$ signifies the angle matrix, where in $D_{ij} = \sum_{j=1}^{n} A_{ij}$ and $\sigma$ represents the ReLU activation function. Then, SAGEConv layer leverages the neighbor node features of $G$ to refine the feature representation of each node. Specifically, for each node $v$, its neighbor node set $N(v)$ is selected, and the mean pooling aggregation method is applied to combine the features of its neighboring nodes. The feature $H_v^{(l)}$ of node $v$ is then concatenated with the aggregated features of its neighbors, and after undergoing a linear transformation and nonlinear activation function, the updated representation $H^{(l+1)}$ of node $v$ is derived. This process is formally expressed as follows:

$$H_v^{(l+1)} = \sigma(W \cdot Concat(H_v^{(l)}, Aggregate(\{H_u^{(l)}, \forall_u \in N(v)\}))) \quad (2)$$

$W$ is the weight matrix, *Concat* concatenates the features of a node with those of its neighbors, and Aggregate is the aggregation operation [66]. The features of the $G$ are processed through a ReLU layer and a Dropout layer, followed by feature extraction and dimensionality reduction using an MLP network and a Rescale layer [67], [68]. Finally, the features output by the two symmetric encoders are concatenated, and their mean is calculated to obtain the final output of VERN.

A key characteristic of the VERN is its use of cross-graph message passing [69]. This means that the two encoders not only process their respective graph features but also exchange information through weight sharing. This mechanism enables the upper encoder to learn more robust feature representations by sharing the weights of the lower encoder. At the same time, the feature-sharing mechanism helps both encoders to learn more general graph features, thus improving the model's generalization ability [70]. When handling 768-dimensional feature vectors, the model also incorporates a skip connection that passes the initial input information to later layers. This not only preserves key information from the original input but also improves gradient propagation, making the model training more stable.

**Experimental setup and implementation details**
The VERN utilizes an internal validation set and employs five-fold cross-validation to evaluate its performance. In this cross-validation method, the internal validation set is divided into five groups; four are used for training and one for validation in each fold, resulting in five optimal prediction models. The VERN is then validated on an internal test set. Most importantly, the model's generalizability is assessed using both single-cohort and multicenter validation sets.

Through grid search, we tuned the learning rate, regularization strength, and batch size and finally determined the current parameter combination. The VERN was trained using the RMSprop optimizer with a learning rate of 0.001 and an alpha weight decay rate of 0.9 [71]. The batch size was set to 1 mainly because the number of patches in each WSI varies, and the sizes of the graph adjacency matrices differ. As a result, it is impossible to stack graphs of different sizes into batches with uniform output. Additionally, the VERN consumes a significant amount of GPU memory during inference, and increasing the batch size could lead to memory overflow. The number of epochs was set to 200, with each epoch iterating through the training data once. To conduct comparative experiments, we implemented SOTA methods for ABMIL, DSMIL, TransMIL, DTFD-MIL and IBMIL [32], [33], [34], [35], [36].

**Visualization and explainability**
The final prediction is obtained by associating the class tag with the input sequence. The encoder output feature of the VERN, namely, the adjacency matrix, generates a one-dimensional vector representing the contribution of each patch.

To mitigate contribution bias, we average the contributions of each patch using the symmetric hybrid encoder outputs, and normalize these contributions to the [0,1] range. Subsequently, we redistribute these contributions to each patch, map each patch back onto the WSI, and present the final results as a heat map for better visual interpretation. Using the contribution scores, we identify the top 9 patches with the highest contributions for further analysis.

**Statistical analysis**
We use AUCROC curve, PRC curve, Accuracy, Precision, Recall, F1-Score, Specificity, AUC as our evaluation indicators. The ROC curve represents the relationship between sensitivity and specificity relationship, the larger the area, the better. The PRC represents the relationship between Precision and Recall, with a larger area indicating better performance. Higher values of Accuracy, Precision, Recall, F1-Score, Specificity, and AUC are better [72], [73].

**Hardware and software**
Data processing is conducted on a Windows 11 host equipped with an Intel i7 13700K processor, 128GB of memory, 60TB of storage, and a 24GB NVIDIA GeForce RTX 4090 graphics card. PyTorch 1.13, Python 3.9, and various other open-source tools and packages are utilized for model development, testing, and training. All models are trained on a GPU.

**Ethical review**
All experiments are approved by the Ethics Committee of the Second Xiangya Hospital, Central South University (Ethics verify file Z0331-01).

## Data availability

Due to privacy protections for patient histopathology image data and patient clinical data, we are unable to make all data public. We provide the labels and features of the public datasets at https://github.com/pengsl-lab/STAS/tree/main#datastes. TCGA and CPTAC diagnostic pathology data can be downloaded at https://portal.gdc.cancer.gov/.

## Code availability

The data, source code, and standalone program of VERN are open accessible at https://github.com/pengsl-lab/STAS. The pretraining codes and VERN are available for academic use at https://github.com/pengsl-lab/STAS/tree/main.

## AUTHOR CONTRIBUTIONS

Liangrui Pan, Qingchun Liang: Conceptualization, Methodology, Software, Visualization, Writing original draft. Wenwu Zeng, Yijun Peng, Yan Li, Yiyi Liang, Zhenyu Zhao: data collection. Jiadi Luo: Visualization, Writing original draft, Supervision. Xiang Wang and Shaoliang Peng: Funding acquisition, Resources, Writing-review & editing, Supervision.

## ACKNOWLEDGMENT

This work was supported by Hunan Province Graduate Research Innovation Project CX20240450; NSFC-FDCT Grants 62361166662; National Key R&D Program of China 2023YFC3503400, 2022YFC3400400; Key R&D Program of Hunan Province 2023GK2004, 2023SK2059, 2023SK2060; Top 10 Technical Key Project in Hunan Province 2023GK1010; Key Technologies R&D Program of Guangdong Province (2023B1111030004 to FFH). Key R&D Program of Hunan Province 2023GK2004, National Natural Science Foundation of China Youth Science Fund Project: 82200019. The Funds of State Key Laboratory of Chemo/Biosensing and Chemometrics, the National Supercomputing Center in Changsha (http://nscc.hnu.edu.cn/), and Peng Cheng Lab. J.D. Luo, X. Wang and S.L. Peng are the corresponding authors for this paper.

**CONFLICT OF INTEREST**

The authors declare that they have no known competing financial interests or personal relationships that could have appeared to influence the work reported in this paper.

**REFERENCES**

[1] M. L. Onozato et al., "Tumor islands in resected early-stage lung adenocarcinomas are associated with unique clinicopathologic and molecular characteristics and worse prognosis," *The American journal of surgical pathology*, vol. 37, no. 2, pp. 287–294, 2013.

[2] K. Kadota et al., "Tumor spread through air spaces is an important pattern of invasion and impacts the frequency and location of recurrences after limited resection for small stage I lung adenocarcinomas," *Journal of Thoracic Oncology*, vol. 10, no. 5, pp. 806–814, 2015.

[3] A. Warth et al., "Prognostic impact of intra-alveolar tumor spread in pulmonary adenocarcinoma," *The American journal of surgical pathology*, vol. 39, no. 6, pp. 793–801, 2015.

[4] M. Chae et al., "Prognostic significance of tumor spread through air spaces in patients with stage IA part-solid lung adenocarcinoma after sublobar resection," *Lung Cancer*, vol. 152, pp. 21–26, Feb. 2021, doi: 10.1016/j.lungcan.2020.12.001.

[5] T. Eguchi et al., "Lobectomy is associated with better outcomes than sublobar resection in spread through air spaces (STAS)-positive T1 lung adenocarcinoma: a propensity score–matched analysis," *Journal of thoracic oncology*, vol. 14, no. 1, pp. 87–98, 2019.

[6] S. Shiono and N. Yanagawa, "Spread through air spaces is a predictive factor of recurrence and a prognostic factor in stage I lung adenocarcinoma," *Interactive cardiovascular and thoracic surgery*, vol. 23, no. 4, pp. 567–572, 2016.

[7] C. Dai et al., "Tumor spread through air spaces affects the recurrence and overall survival in patients with lung adenocarcinoma> 2 to 3 cm," *Journal of thoracic oncology*, vol. 12, no. 7, pp. 1052–1060, 2017.

[8] E. Yi, M. Bae, S. Cho, J. Chung, S. Jheon, and K. Kim, "Pathological prognostic factors of recurrence in early stage lung adenocarcinoma," *ANZ journal of surgery*, vol. 88, no. 4, pp. 327–331, 2018.

[9] Y. Terada et al., "Spread through air spaces is an independent predictor of recurrence in stage III (N2) lung adenocarcinoma," *Interactive cardiovascular and thoracic surgery*, vol. 29, no. 3, pp. 442–448, 2019.

[10] H. A. Khalil et al., "Analysis of recurrence in lung adenocarcinoma with spread through air spaces," *The Journal of Thoracic and Cardiovascular Surgery*, vol. 166, no. 5, pp. 1317–1328, 2023.

[11] T. Xia, Q. Yuan, and S. Xing, "STAS: New explorations and challenges for thoracic surgeons," *Clinical and Translational Oncology*, pp. 1–11, 2024.

[12] W. D. Travis et al., "The International Association for the Study of Lung Cancer (IASLC) Staging Project for Lung Cancer: Recommendation to Introduce Spread Through Air Spaces as a Histologic Descriptor in the Ninth Edition of the TNM Classification of Lung Cancer. Analysis of 4061 Pathologic Stage I NSCLC," *Journal of Thoracic Oncology*, vol. 19, no. 7, pp. 1028–1051, Jul. 2024, doi: 10.1016/j.jtho.2024.03.015.

[13] K. Xu et al., "Prognostic significance of limited resection in pathologic stage I lung adenocarcinoma with spread through air spaces," *Journal of Thoracic Disease*, vol. 15, no. 9, p. 4795, 2023.

[14] E. Balaur et al., "Colorimetric histology using plasmonically active microscope slides," *Nature*, vol. 598, no. 7879, pp. 65–71, Oct. 2021, doi: 10.1038/s41586-021-03835-2.

[15] A. Hekler et al., "Deep learning outperformed 11 pathologists in the classification of histopathological melanoma images," *European Journal of Cancer*, vol. 118, pp. 91–96, Sep. 2019, doi: 10.1016/j.ejca.2019.06.012.

[16] "Pathology Visions 2020: Through the Prism of Innovation," *Journal of Pathology Informatics*, vol. 12, no. 1, p. 37, Jan. 2021, doi: 10.4103/2153-3539.326643.

[17] W. K. Funkhouser, "Pathology: the clinical description of human disease," in *Essential Concepts in Molecular Pathology*, Elsevier, 2020, pp. 177–190.

[18] H. Kawasaki et al., "The NanoSuit method: a novel histological approach for examining paraffin sections in a nondestructive manner by correlative light and electron microscopy," *Laboratory Investigation*, vol. 100, no. 1, pp. 161–173, Jan. 2020, doi: 10.1038/s41374-019-0309-7.

[19] Y. Ding et al., "The value of frozen section diagnosis of tumor spread through air spaces in small-sized (≤2 cm) non-small cell lung cancer," *World Journal of Surgical Oncology*, vol. 21, no. 1, p. 195, Jul. 2023, doi: 10.1186/s12957-023-03092-9.

[20] H. Cao et al., "Prediction of Spread Through Air Spaces (STAS) By Intraoperative Frozen Section for Patients with cT1N0M0 Invasive Lung Adenocarcinoma: A Multi-Center Observational Study (ECTOP-1016)," *Annals of Surgery*, pp. 10–1097.

[21] J. A. Villalba et al., "Accuracy and reproducibility of intraoperative assessment on tumor spread through air spaces in stage 1 lung adenocarcinomas," *Journal of Thoracic Oncology*, vol. 16, no. 4, pp. 619–629, 2021.

[22] F. Zhou et al., "Assessment of the feasibility of frozen sections for the detection of spread through air spaces (STAS) in pulmonary adenocarcinoma," *Modern Pathology*, vol. 35, no. 2, pp. 210–217, 2022.

[23] A. Hosny, C. Parmar, J. Quackenbush, L. H. Schwartz, and H. J. W. L. Aerts, "Artificial intelligence in radiology," *Nat Rev Cancer*, vol. 18, no. 8, pp. 500–510, Aug. 2018, doi: 10.1038/s41568-018-0016-5.

[24] K. Choudhary et al., "Recent advances and applications of deep learning methods in materials science," *npj Computational Materials*, vol. 8, no. 1, p. 59, Apr. 2022, doi: 10.1038/s41524-022-00734-6.

[25] N. Marini, S. Otálora, H. Müller, and M. Atzori, "Semi-supervised training of deep convolutional neural networks with heterogeneous data and few local annotations: An experiment on prostate histopathology image classification," *Medical Image Analysis*, vol. 73, p. 102165, Oct. 2021, doi: 10.1016/j.media.2021.102165.

[26] M. Chen et al., "Classification and mutation prediction based on histopathology H&E images in liver cancer using deep learning," *npj Precis. Onc.*, vol. 4, no. 1, p. 14, Dec. 2020, doi: 10.1038/s41698-020-0120-3.

[27] H. Cai et al., "MIST: multiple instance learning network based on Swin Transformer for whole slide image classification of colorectal adenomas," *The Journal of Pathology*, vol. 259, no. 2, pp. 125–135, 2023.

[28] Z. He et al., "Deconv-transformer (DecT): A histopathological image classification model for breast cancer based on color deconvolution and transformer architecture," *Information Sciences*, vol. 608, pp. 1093–1112, Aug. 2022, doi:

[29] H. Failmezger, S. Muralidhar, A. Rullan, C. E. de Andrea, E. Sahai, and Y. Yuan, "Topological Tumor Graphs: A Graph-Based Spatial Model to Infer Stromal Recruitment for Immunosuppression in Melanoma Histology," *Cancer Research*, vol. 80, no. 5, pp. 1199–1209, Mar. 2020, doi: 10.1158/0008-5472.CAN-19-2268.

[30] Z. Wu *et al.*, "Graph deep learning for the characterization of tumour microenvironments from spatial protein profiles in tissue specimens," *Nature Biomedical Engineering*, vol. 6, no. 12, pp. 1435–1448, Dec. 2022, doi: 10.1038/s41551-022-00951-w.

[31] B. He *et al.*, "Integrating spatial gene expression and breast tumour morphology via deep learning," *Nature Biomedical Engineering*, vol. 4, no. 8, pp. 827–834, Aug. 2020, doi: 10.1038/s41551-020-0578-x.

[32] M. Ilse, J. Tomczak, and M. Welling, "Attention-based deep multiple instance learning," presented at the International conference on machine learning, PMLR, 2018, pp. 2127–2136.

[33] B. Li, Y. Li, and K. W. Eliceiri, "Dual-stream multiple instance learning network for whole slide image classification with self-supervised contrastive learning," presented at the Proceedings of the IEEE/CVF conference on computer vision and pattern recognition, 2021, pp. 14318–14328.

[34] Z. Shao, H. Bian, Y. Chen, Y. Wang, J. Zhang, and X. Ji, "Transmil: Transformer based correlated multiple instance learning for whole slide image classification," *Advances in Neural Information Processing Systems*, vol. 34, pp. 2136–2147, 2021.

[35] H. Zhang *et al.*, "DTFD-MIL: Double-Tier Feature Distillation Multiple Instance Learning for Histopathology Whole Slide Image Classification," in *2022 IEEE/CVF Conference on Computer Vision and Pattern Recognition (CVPR)*, New Orleans, LA, USA: IEEE, Jun. 2022, pp. 18780–18790. doi: 10.1109/CVPR52688.2022.01824.

[36] T. Lin, Z. Yu, H. Hu, Y. Xu, and C. W. Chen, "Interventional Bag Multi-Instance Learning On Whole-Slide Pathological Images," in *2023 IEEE/CVF Conference on Computer Vision and Pattern Recognition (CVPR)*, 2023, pp. 19830–19839. doi: 10.1109/CVPR52729.2023.01899.

[37] I. Heckenbach *et al.*, "Nuclear morphology is a deep learning biomarker of cellular senescence," *Nature Aging*, vol. 2, no. 8, pp. 742–755, Aug. 2022, doi: 10.1038/s43587-022-00263-3.

[38] F. M. Shiri, T. Perumal, N. Mustapha, and R. Mohamed, "A comprehensive overview and comparative analysis on deep learning models: CNN, RNN, LSTM, GRU," *arXiv preprint arXiv:2305.17473*, 2023.

[39] J. Metovic *et al.*, "Gross specimen handling procedures do not impact the occurrence of spread through air spaces (STAS) in lung cancer," *The American journal of surgical pathology*, vol. 45, no. 2, pp. 215–222, 2021.

[40] J. L. G. Blaauwgeers, "Histopathological aspects of resected non-small cell lung cancer, with emphasis on spread through air spaces and collapsed adenocarcinoma in situ," 2023.

[41] H. Dong, H. Ma, Z. Du, Z. Zhou, H. Yang, and Z. Wang, "Graph learning considering dynamic structure and random structure," *Journal of King Saud University-Computer and Information Sciences*, vol. 35, no. 7, p. 101633, 2023.

[42] J. Petersson and R. W. Glenny, "Gas exchange and ventilation–perfusion relationships in the lung," *European Respiratory Journal*, vol. 44, no. 4, pp. 1023–1041, 2014.

[43] Y. Zhang *et al.*, "Global variations in lung cancer incidence by histological subtype in 2020: a population-based study," *The Lancet Oncology*, vol. 24, no. 11, pp. 1206–1218, 2023.

[44] Z. Chen *et al.*, "Prognostic impact of tumor spread through air spaces for T2aN0 stage IB non-small cell lung cancer," *Cancer Medicine*, vol. 12, no. 14, pp. 15246–15255, 2023.

[45] S.-Y. Hu *et al.*, "Correlation of tumor spread through air spaces and clinicopathological characteristics in surgically resected lung adenocarcinomas," *Lung Cancer*, vol. 126, pp. 189–193, 2018.

[46] Y. Liu *et al.*, "Relationship between MTA1 and spread through air space and their joint influence on prognosis of patients with stage I-III lung adenocarcinoma," *Lung cancer*, vol. 124, pp. 211–218, 2018.

[47] J. S. Lee, E. K. Kim, M. Kim, and H. S. Shim, "Genetic and clinicopathologic characteristics of lung adenocarcinoma with tumor spread through air spaces," *Lung Cancer*, vol. 123, pp. 121–126, 2018.

[48] N. Singh *et al.*, "p53 immunohistochemistry is an accurate surrogate for TP53 mutational analysis in endometrial carcinoma biopsies," *The Journal of Pathology*, vol. 250, no. 3, pp. 336–345, 2020.

[49] N. Matsumoto *et al.*, "Correlative assessment of p53 immunostaining patterns and TP53 mutation status by next-generation sequencing in high-grade endometrial carcinomas," *International Journal of Gynecological Pathology*, vol. 42, no. 6, pp. 567–575, 2023.

[50] K. Wakuda *et al.*, "Concordance of ALK fusion gene-rearrangement between immunohistochemistry and next-generation sequencing," *International Journal of Clinical Oncology*, vol. 29, no. 2, pp. 96–102, 2024.

[51] Y. Tian *et al.*, "Integration of clinicopathological and mutational data offers insight into lung cancer with tumor spread through air spaces," *Annals of Translational Medicine*, vol. 9, no. 12, 2021.

[52] J. Wang, Y. Yao, D. Tang, and W. Gao, "An individualized nomogram for predicting and validating spread through air space (STAS) in surgically resected lung adenocarcinoma: a single center retrospective analysis," *Journal of Cardiothoracic Surgery*, vol. 18, no. 1, p. 337, 2023.

[53] M. Jia, S. Yu, J. Yu, Y. Li, H. Gao, and P.-L. Sun, "Comprehensive analysis of spread through air spaces in lung adenocarcinoma and squamous cell carcinoma using the 8th edition AJCC/UICC staging system," *BMC cancer*, vol. 20, pp. 1–11, 2020.

[54] S. Altinay *et al.*, "Spread through air spaces (STAS) is a predictor of poor outcome in atypical carcinoids of the lung," *Virchows Archiv*, vol. 475, pp. 325–334, 2019.

[55] M. Spiliotaki *et al.*, "Dynamic monitoring of PD-L1 and Ki67 in circulating tumor cells of metastatic non-small cell lung cancer patients treated with pembrolizumab," *Molecular Oncology*, vol. 17, no. 5, pp. 792–809, 2023.

[56] L. Huang, R. Luo, X. Liu, and X. Hao, "Spectral imaging with deep learning," *Light: Science & Applications*, vol. 11, no. 1, p. 61, 2022.

[57] A. Riasatian *et al.*, "Fine-Tuning and training of densenet for histopathology image representation using TCGA diagnostic slides," *Medical Image Analysis*, vol. 70, p. 102032, 2021, doi: https://doi.org/10.1016/j.media.2021.102032.

[58] X. Wang *et al.*, "Transformer-based unsupervised contrastive learning for histopathological image classification," *Medical Image Analysis*, vol. 81, p. 102559, Oct. 2022, doi: 10.1016/j.media.2022.102559.

[59] G. Huang, Z. Liu, L. Van Der Maaten, and K. Q. Weinberger, "Densely Connected Convolutional Networks," in *2017 IEEE Conference on Computer Vision and Pattern Recognition (CVPR)*, Honolulu, HI: IEEE, Jul. 2017, pp. 2261–2269. doi:

10.1109/CVPR.2017.243.

[60] Z. Liu *et al.*, "Swin transformer: Hierarchical vision transformer using shifted windows," presented at the Proceedings of the IEEE/CVF International Conference on Computer Vision, 2021, pp. 10012–10022.

[61] J. M. Keller, M. R. Gray, and J. A. Givens, "A fuzzy k-nearest neighbor algorithm," *IEEE transactions on systems, man, and cybernetics*, no. 4, pp. 580–585, 1985.

[62] R. J. Chen *et al.*, "Pathomic Fusion: An Integrated Framework for Fusing Histopathology and Genomic Features for Cancer Diagnosis and Prognosis," *IEEE Transactions on Medical Imaging*, vol. 41, no. 4, pp. 757–770, Apr. 2022, doi: 10.1109/TMI.2020.3021387.

[63] T. N. Kipf and M. Welling, "Semi-supervised classification with graph convolutional networks," *arXiv preprint arXiv:1609.02907*, 2016.

[64] W. L. Hamilton, R. Ying, and J. Leskovec, "Inductive Representation Learning on Large Graphs," in *Proceedings of the 31st International Conference on Neural Information Processing Systems*, in NIPS'17. Red Hook, NY, USA: Curran Associates Inc., 2017, pp. 1025–1035.

[65] A. F. Agarap, "Deep Learning using Rectified Linear Units (ReLU)," *arXiv:1803.08375 [cs, stat]*, Feb. 2019, Accessed: Mar. 15, 2020. [Online]. Available: http://arxiv.org/abs/1803.08375

[66] J. Chang, L. Wang, G. Meng, Q. Zhang, S. Xiang, and C. Pan, "Local-aggregation graph networks," *IEEE Transactions on Pattern Analysis and Machine Intelligence*, vol. 42, no. 11, pp. 2874–2886, 2019.

[67] H. Taud and J.-F. Mas, "Multilayer perceptron (MLP)," *Geomatic approaches for modeling land change scenarios*, pp. 451–455, 2018.

[68] V. Papyan, X. Han, and D. L. Donoho, "Prevalence of neural collapse during the terminal phase of deep learning training," *Proceedings of the National Academy of Sciences*, vol. 117, no. 40, pp. 24652–24663, 2020.

[69] H. Wang, P. Cao, J. Wang, and O. R. Zaiane, "Uctransnet: rethinking the skip connections in u-net from a channel-wise perspective with transformer," presented at the Proceedings of the AAAI conference on artificial intelligence, 2022, pp. 2441–2449.

[70] S. F. Ahmed *et al.*, "Deep learning modelling techniques: current progress, applications, advantages, and challenges," *Artificial Intelligence Review*, vol. 56, no. 11, pp. 13521–13617, 2023.

[71] F. Zou, L. Shen, Z. Jie, W. Zhang, and W. Liu, "A sufficient condition for convergences of adam and rmsprop," presented at the Proceedings of the IEEE/CVF Conference on computer vision and pattern recognition, 2019, pp. 11127–11135.

[72] D. Chicco and G. Jurman, "The advantages of the Matthews correlation coefficient (MCC) over F1 score and accuracy in binary classification evaluation," *BMC genomics*, vol. 21, pp. 1–13, 2020.

[73] T. B. Alakus and I. Turkoglu, "Comparison of deep learning approaches to predict COVID-19 infection," *Chaos, Solitons & Fractals*, vol. 140, p. 110120, 2020.